\documentclass[journal=jctcce,%
               manuscript=article,%
               layout=twocolumn]{achemso}

\usepackage[T1]{fontenc} 
\usepackage{amssymb}
\usepackage{mathtools}
\usepackage{siunitx}
\usepackage{hyperref}
\usepackage{microtype}

\newcommand{\SIFigMI}{S1}
\newcommand{\SIFigCluster}{S2}
\newcommand{\SIFigCorr}{S3}
\newcommand{\SIFigLys}{S4}
\newcommand{\SIFigVillin}{S5}

\author{Georg Diez}
\altaffiliation{G. Diez and D. Nagel contributed equally to this work.}
\author{Daniel Nagel}
\altaffiliation{G. Diez and D. Nagel contributed equally to this work.}
\author{Gerhard Stock}
\affiliation{Biomolecular Dynamics, Institute of Physics,
Albert-Ludwigs-Universit\"at, 79104 Freiburg, Germany}
\email{stock@physik.uni-freiburg.de}

\title{Correlation-based feature selection to identify functional
dynamics in proteins}

\begin{document}


\begin{abstract}
  To interpret molecular dynamics simulations of biomolecular systems,
  systematic dimensionality reduction methods are commonly
  employed. Among others, this includes principal component analysis
  (PCA) and time-lagged independent component analysis (TICA), which aim
  to maximize the variance and the timescale of the first components,
  respectively. A crucial first step of such an analysis is the
  identification of suitable and relevant input coordinates (the
  so-called features), such as backbone dihedral angles and interresidue
  distances. As typically only a small subset of those coordinates is
  involved in a specific biomolecular process, it is important to
  discard the remaining uncorrelated motions or weakly correlated noise
  coordinates. This is because they may exhibit large amplitudes or long
  timescales and therefore will be erroneously be considered important
  by PCA and TICA, respectively. To discriminate collective motions
  underlying functional dynamics from uncorrelated motions, the
  correlation matrix of the input coordinates is block-diagonalized by a
  clustering method. This strategy avoids possible bias due to presumed
  functional observables and conformational states or variation
  principles that maximize variance or timescales.
  Considering several linear and nonlinear
  correlation measures and various clustering algorithms, it is shown that
  the combination of linear correlation and the Leiden community
  detection algorithm yields excellent results for all considered model
  systems. These include the functional motion of T4 lysozyme to
  demonstrate the successful identification of collective motion, as
  well as the folding of villin headpiece to highlight the physical
  interpretation of the correlated motions in terms of a functional
  mechanism.
\end{abstract}

\section{Introduction}

Molecular dynamics (MD) simulation is a versatile and widely used
approach to study the spatiotemporal dynamics of biomolecular
systems.\cite{Berendsen07} Since it is neither possible nor desirable
to follow the motion of a complex molecule along its $3N$ atomic
coordinates, a low-dimensional representation of the dynamics is
required, which explains the mechanism and the underlying structural
rearrangements of some biomolecular
process. \cite{Bolhuis00,Faradjian04,Best05,Krivov08a, E10,McGibbon17}
To this end, a number of efficient and systematic strategies of
dimensionality reduction have been developed.\cite{Jolliffe02,Lee07}
Popular examples include principal component analysis\cite{Amadei93}
(PCA) which represents a linear transformation to coordinates that
maximize the variance of the first components, and time-lagged
independent component analysis\cite{Perez-Hernandez13} (TICA) which
aims to maximize the timescales of the first components. Moreover a
variety of nonlinear techniques as well as a rapidly increasing number
of machine-learning empowered approaches have been proposed, see,
e.g., Refs.~\citenum{Rohrdanz13,Wang20,Glielmo21} for reviews.

In this work, we are concerned with the crucial initial step in
biomolecular dimensionality reduction, that is, the identification of
suitable and relevant input coordinates for the
analysis.\cite{Sittel18} First the type of coordinates needs to be
chosen. Due to inevitable mixing of overall rotation and internal
motion, Cartesian coordinates are in general not suited for
dimensionality reduction.\cite{note1,Sittel14} Internal coordinates
such as dihedral angles and interatomic distances, on the other hand,
are by definition not plagued by this problem and also represent a
natural choice, since the molecular force field is given in terms of
internal coordinates. While $(\phi,\,\psi)$ backbone dihedral angles
have been shown to accurately describe the conformation of secondary
structures such as $\alpha$-helices and $\beta$-sheets,
\cite{Altis08,Fenwick14} interresidue distances appear to be well
suited to characterize the overall structure of a
protein.\cite{Laetzer08,Hori09,Kalgin13} A drawback of using interresidue
distances is that their number scales quadratically with the number of
residues. To avoid this overrepresentation, it has been suggested to
restrict the analysis on distances reflecting interresidue contacts
such as hydrogen bonds, salt bridges, and hydrophobic
contacts.\cite{Ernst15} In this way, we focus on near order effects
rather than long distances, because the latter can be typically
understood as a consequence of contact changes.

Irrespective of the type of input coordinates, we may expect that only
a subset of them will be involved in a specific biomolecular
process. A general approach to identify these relevant coordinates of
a process is to consider their mutual relation, as quantified by some
correlation measure.\cite{Lange06,Lange08a} For example, functional mode
analysis\cite{Hub09,Krivobokova12} aims to identify
molecular coordinates that significantly correlate with some
pre-defined functional observable of the system, reflecting, e.g., the
transition from an active to an inactive conformational state. This
strategy excludes coordinates that do not change during the functional
motion (e.g., stable contacts), as well as coordinates that change
randomly such as wildly dangling terminal residues. Exclusion of the
latter is particularly important if we intend to subsequently perform
a PCA. Maximizing the variance of the first principal components, PCA
would rate this terminal large-amplitude motion as important, although
it is generally irrelevant for protein function. Maximizing the
timescales of the first components, a similar problem occurs for TICA
if the coordinates contain slow but irrelevant motions.\cite{Husic19}
For example, when performing TICA on $(\phi,\,\psi)$ dihedral angles
to describe the folding of the helical protein HP35,\cite{Sittel18}
it correctly identifies transitions between right- and left-handed
helices as slowest process.\cite{Nagel20} Since the left-handed helices
are hardly populated, however, this motion is not relevant for the
folding process and the corresponding deceptive coordinates should
therefore be discarded in the further analysis.\cite{Sittel18}

Here we wish to identify coordinates, often called ``features'', which
describe motions that are involved in a specific process, and thus
discard the remaining coordinates from the analysis, describing other
processes and random motion or noise. Due to the reduction of noise
and its lower dimensionality, the resulting feature space
considerably facilitates the subsequent analysis and may even lead to
a straightforward interpretation of the considered process. As we aim
to perform the feature selection in an unbiased manner, we avoid
variation principles that maximize variance or
timescales.\cite{Scherer19} Moreover, we do not want to invoke
previously known functional observables as in functional mode analysis
\cite{Hub09} or refer to pre-defined metastable conformational states
as in a recently proposed supervised machine learning
scheme.\cite{Brandt18} Rather, we follow the strategy pursued by Tiwary
and coworkers\cite{Ravindra20} and first calculate the correlation
matrix of all input coordinates to establish their interrelation, and
then block-diagonalize this matrix to unravel which set of coordinates
is associated with the process under consideration.
A somewhat related approach is followed by various groups that aim to
construct a dynamical network of protein residues in order to map out
pathways of allosteric communication.\cite{Sethi09, McClendon09,
  Bowman12, Dokholyan16}
Representing the
features as data points in a similarity space where the features are
arranged according to their proximity, the latter step can be viewed
as a straightforward clustering task which identifies groups of
nearby data points. Numerous options exist to achieve this
clustering,\cite{Saxena17} including hierarchical methods (such as
complete linkage clustering\cite{Voorhees86}), geometrical approaches
(such as $k$-medoids\cite{Kaufman90}) or graph-based community
detection methods.\cite{Fortunato10,Newman10}

In this work we introduce the Python package MoSAIC (``Molecular
Systems Automated Identification of Cooperativity''), which
automatically detects collective motions in MD simulation data,
identifies uncorrelated coordinates as noise, and hence provides a
detailed picture of the key coordinates driving a conformational
change in a biomolecular system. Considering several linear and
nonlinear correlation measures and various clustering algorithms, we
show that the combination of linear correlation and the Leiden
community detection algorithm \cite{Traag19} yields excellent
results for all considered model systems.
These include a model correlation matrix with known ground truth to
test various clustering approaches, the functional motion of T4
lysozyme \cite{Ernst17} to demonstrate the successful identification
of collective motion, as well as the folding of villin headpiece
\cite{Piana12} to highlight the physical interpretation of the
resulting clusters in terms of a functional mechanism.

\section{Correlation Measures}

In order to establish a suitable similarity measure to identify
collective motion in proteins, we introduce various definitions of
the correlation and study their performance for our set of model problems.

\subsection{Linear and nonlinear correlation}

Given two random variables $X$ and $Y$, the linear correlation $\rho$
is defined as the Pearson correlation coefficient
\begin{align}
    \rho(X,Y) = \frac{\langle (X-\langle X \rangle)(Y- \langle Y
  \rangle)\rangle}{\sigma_X \sigma_Y}\,,
    \label{eq:Pearson}
\end{align}
where $\langle \ldots \rangle$ denotes the statistical average and
$\sigma$ is the standard deviation. $\rho$ ranges from $-1$ to $1$,
where a value of $0$ implies no linear correlation at all while values
of $\pm 1$ corresponds to a perfect linear relationship between $X$
and $Y$.  In that case, one variable can be fully described by the
other variable, and may therefore be discarded if we are interested in
dimensionality reduction.
By definition, the Pearson coefficient considers only the first two
moments of the underlying distribution, which---strictly speaking---is
adequate only for normally distributed data.  From a Physics point of
view, this translates into a quadratic energy landscape associated with
a linear force.\cite{Berendsen07}

To go beyond the linear regime, we consider the mutual information
$I(X,Y)$ of two variables, which can be defined as \cite{Cover06}
\begin{align}
    I(X,Y) &= H(X)-H(X|Y)\,, \label{eq:mutualinformation_entropy} \\
    H(X) &= - \sum_{x \in X} p(x) \log p(x) \,,
\end{align}
where $H(X)$ is the information entropy that measures the uncertainty
of a discrete random variable $X$ with samples $x$, $H(X|Y)$ is the
corresponding conditional entropy, and $p(x)$
denotes the probability distribution of $X$. Since $H(X)$ can be
viewed as the uncertainty about $X$, $H(X|Y)$ may be interpreted as
the uncertainty about $X$ remaining after knowing $Y$, i.e., it
indicates what $Y$ does not tell us about $X$.
Consequently, mutual information can be viewed as the reduction in
uncertainty of $X$ due to the knowledge of $Y$, thus offering a
theoretically well justified measure of how similar two random
variables are.
Alternatively, the mutual information can be defined as measure of
statistical independence of variables $X$ and $Y$, i.e., the
discrepancy between the joint distribution $p(x,y)$ and the product of
the marginal distributions $p(x)p(y)$. Employing the
Kullback-Leibler divergence $D_{\text{KL}} \left[ p_1||p_2 \right]$ as
a commonly used measure for the dissimilarity of two probability
distributions $p_1$ and $p_2$, we obtain
\begin{align}
    I(X,Y)
        &= D_{\text{KL}} \left[ p(x,y)||p(x)p(y) \right]\nonumber\\
        &= \sum_{x\in X} \sum_{y\in Y} p(x,y) \log_2 \left[ \frac{p(x,y)}{p(x)p(y)}\right]\,.
    \label{eq:mutualinfo}
\end{align}

Being defined via probability distributions that contain all
statistical moments, the mutual information is not restricted to
linear correlation as the Pearson correlation coefficient in
Eq.~\eqref{eq:Pearson}. On the other hand, the mutual information
is not bound to $[-1, 1]$ but ranges from $[0, \infty)$, which is
a drawback, as it is not obvious which range of the mutual
information can be considered as high or low correlation.
To this end, several options are at hand to normalize the mutual information.
\cite{Vinh10} Employing the inequality
\begin{equation}
    I(X,Y) \leq \sqrt{H(X)H(Y)} \leq H(X,Y) \,,
    \label{eq:mutual_information_bounds}
\end{equation}
we may adopt the geometric mean of marginal entropies
$\sqrt{H(X)H(Y)}$ or the joint
entropy $H(X,Y)$ to define the normalized quantities
\begin{align}
    I_\text{geom.}(X, Y) &= \frac{I(X,Y)}{\sqrt{H(X)H(Y)}}\,,\\
    I_\text{joint}(X, Y) &= \frac{I(X,Y)}{H(X, Y)}\,.
    \label{eq:normalized_MI}
\end{align}
Alternatively, we can use the Jensen-Shannon divergence
$D_\textsc{js}$ as a similarity measure of $p(x,y)$ and $p(x)p(y)$,
which by design it is bound to $[0, 1]$. This leads to
\begin{align}
    & I_\textsc{js}(X, Y)
	= D_\textsc{js} \left[ p(x,y)||p(x)p(y) \right] \\
	& \quad = \frac{1}{2}D_\textsc{kl}\left[p(x,y)||M\right] +
	   \frac{1}{2}D_\textsc{kl}\left[p(x)p(y)||M\right]
\end{align}
with $M = \frac{1}{2}\left[ p(x,y) + p(x)p(y)\right]$.
As a final option we mention the formulation of Gel'fand and Yaglom,
\cite{Gelfand59} who showed that in the special case that the
joint probability distribution $p(x,y)$ is a bivariate normal
distribution, the mutual information $I$ can be deduced exactly from
linear correlation $\rho$ via
\begin{align}
    I(X,Y) &= - \frac{1}{2} \log [1-\rho(X,Y)^2]\,.
    \label{eq:GelfandYaglom}
\end{align}
The identity can be used to normalize the mutual
information by solving for $\rho$ and defining
\begin{align}
    I_\textsc{gy}(X,Y) \equiv \rho(X,Y)
      &= \sqrt{1 - \exp\left[-2 I(X,Y)\right]} \,,
    \label{eq:MI_GY}
\end{align}
which corresponds to a transformation of the mutual information to a
linear correlation.

While the linear correlation in Eq.~\eqref{eq:Pearson} is readily
calculated, the computation of the mutual information is
significantly more involved, because it requires the estimation of (at
least) two-dimensional probability distributions.  Apart from a simple
histogram ansatz, which converges slowly  (with respect to the sample size)
and is not very robust, \cite{Freedman81} we may employ a kernel density
estimation, which is numerically expensive and requires the selection
of a bandwidth parameter,\cite{Heidenreich13} or a non-parametric
$k$-nearest neighbor ($k$-nn) density estimator,\cite{Kraskov04} which is
computationally expensive as well.
Comparing the three methods, kernel density and $k$-nn estimators are
found to converge faster (with respect to the sample size) than the
histogram estimator, but also require factors 100 and 10 more CPU time,
respectively (see Fig.~\SIFigMI a,b). Compared to calculation of the
linear correlation, the reliable estimation of the mutual information
requires at least a factor $\sim 10^3$ more computation time
(Fig.~\SIFigMI c).

\subsection{Choice of similarity measure}

\begin{figure}[h!]
\includegraphics{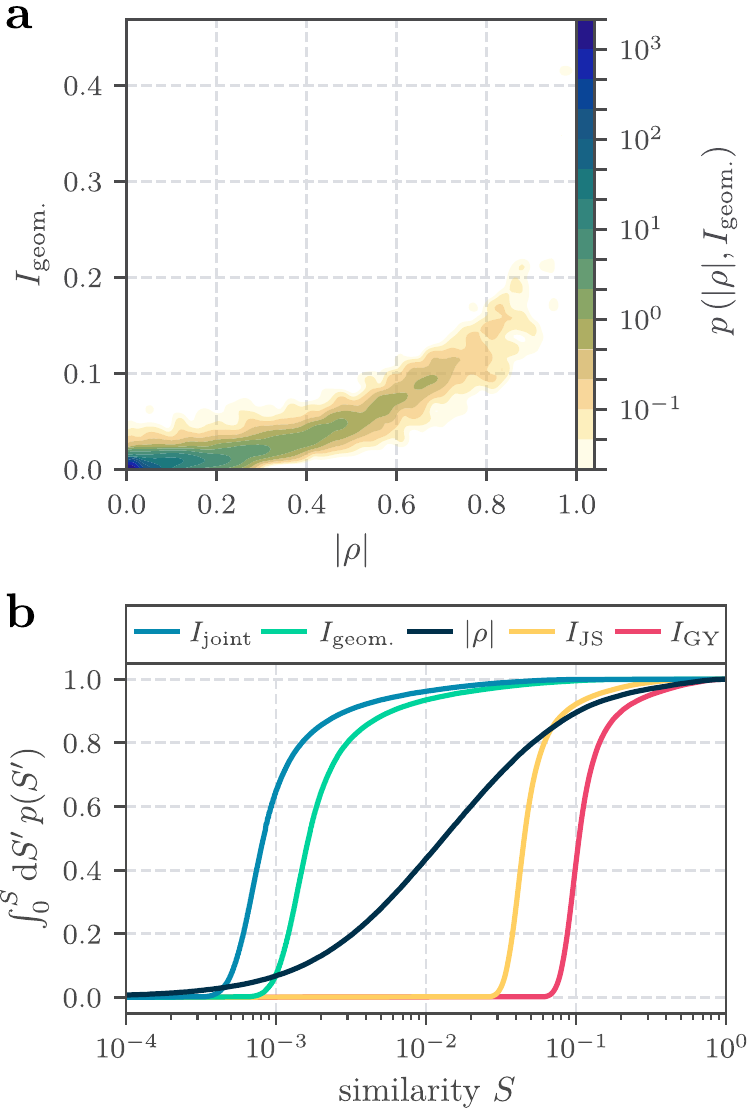}
\caption{Comparison of linear and nonlinear correlation measures,
  obtained from MD data of all protein systems.
(a) Mutual information normalized by the geometric mean, $I_{\text{geom.}}$,
compared to the absolute Pearson coefficient $|\rho|$.
(b) Cumulative probability distribution of various similarity measures.
\label{fig:corr}}
\end{figure}

To assess the performance of the above introduced correlation
measures, we calculated for each measure correlation matrices obtained
from the MD simulations of our model proteins HP35 and T4L (see
Sec.~\ref{sec:appl}). As the results are quite similar for both systems
(Fig.~\SIFigCorr), in the following we combined these data to facilitate the
discussion. To get a first impression of the overall difference
between linear and nonlinear correlations, Fig.~\ref{fig:corr}a
compares the results obtained for the absolute Pearson coefficient
$|\rho|$ and $I_{\text{geom.}}$, the mutual information normalized by
its tightest bound according to Eq.~\eqref{eq:mutual_information_bounds}.
Remarkably, the resulting contour plot reveals a clear relation
between the two correlation measures. Quite similar results are found
for the other normalizations of the mutual information (Fig.~\SIFigCorr).
This appears to indicate that the nonlinear measures do not
contain significant additional information compared to the linear
correlation.

While the values for $|\rho|$ cover the full range from $0$ to $1$,
the values for $I_\text{geom.}$ hardly exceed values of $0.2$.  To
further illustrate this effect, Fig.~\ref{fig:corr}b depicts the
cumulative probability distribution of $I_\text{geom.}$, showing that
more than \SI{90}{\%} of the taken values lie below
$0.01$. Considering the other nonlinear measures, we find that
$I_\text{joint}$ behaves quite similar, while $I_\textsc{JS}$ and
$I_\textsc{GY}$ adopt mainly high values and hardly show values below
$I_\text{JS}=0.03$ or $I_\textsc{GY}=0.07$, respectively.  That is,
although all nonlinear measures are based on the same mutual
information [Eq.~\eqref{eq:mutualinfo}], we hardly find an overlap
between the probability distributions of the various normalized
quantities. The linear correlation $|\rho|$, on the other hand, is
found to uniformly account for small and high correlations.
Combined with the result of Fig.~\ref{fig:corr}a that the mutual
information does not seem to provide essential new information and the
considerably higher ($\gtrsim 10^3$ times) numerical effort
(Fig.~\SIFigMI c), the above findings clearly suggest the simple and
well-established linear Pearson coefficient as suitable similarity
measure.

As a note of caution, we mention that the Pearson coefficient $\rho$
may have serious flaws, in particular, if vector-valued data are
considered.\cite{Matejka17} A well-known example is data
lying on a circle centered around the origin, for which $\rho$ is zero
despite the perfectly correlated data. Restricting ourselves to
scalar data, on the other hand, the Pearson coefficient is known to
capture the overall correlation quite well.\cite{Lange06} Collinear
data in particular includes distances, but also periodic variables (such as
angles and dihedral angles) if an appropriate transformation to
linear variables is applied.\cite{Altis07,Sittel17} That is, the
linear correlation measure should be sufficient for internal molecular
coordinates, but may be problematic for Cartesian
coordinates.\cite{McClendon09}
Moreover we note that the considerable similarity between linear and
nonlinear correlation is somewhat surprising, because the Pearson
coefficient considers only the first two moments of the underlying
distribution [Eq.~\eqref{eq:Pearson}] and the probability
distributions of the employed MD data are not necessarily normally
distributed. For example, we consider contact distances for HP35 and
T4L, which typically reveal a prominent peak at small distances
(reflecting the bound state), and a flat distribution at large
distances indicating unbound conformations. Nevertheless, the
resulting differences between linear and nonlinear correlations were
found to be minor.

%
%
\section{Community detection of collective motion}

Having established a similarity measure to quantify the correlation
between coordinates, we aim to block-diagonalize the resulting
correlation matrix in order to identify the coordinates involved in
some cooperative motion. As discussed in the Introduction, numerous
approaches and algorithms exist for this type of clustering
task. \cite{Saxena17} For the specific application (complex motions of
biomolecules) considered here, we show below that the Leiden community
detection algorithm developed by Traag et al.\cite{Traag19} appears
favorable.

For the further discussion, though, it is instructive to first briefly
introduce two standard methods, $k$-medoids\cite{Kaufman90}
and complete linkage clustering.\cite{Voorhees86} $k$-medoids is a
classical partition technique and bears many similarities with the
popular $k$-means algorithm.\cite{Lloyd82} Given a set of
data points and dissimilarity among them (which we can directly
compute as $1 - |\rho|$), the approach initializes $k$ data points as
medoids and assigns each data point to the nearest medoid. Then
the medoids are greedily optimized (i.e., relying on locally optimal
decisions) until the dissimilarity of all points belonging to a cluster
and the designated medoid (i.e., the center of the cluster) is minimized.
The number of clusters $k$ must be chosen a priori and therefore does
not represent an intrinsic property of the system.

Complete linkage clustering is an agglomerative hierarchical
clustering method which combines data points sequentially into larger
clusters.\cite{Voorhees86} While initially each data point represents
its own cluster, at each step of the algorithm the clusters separated
by the shortest distance are merged. In case the clusters contain more
than a single data point, the distance between clusters is regarded as
the maximum distance between any two elements of the two clusters.
Repeating this step multiple times and merging the closest clusters
together in each step, we eventually end up with a single cluster. To
illustrates the sequence of merging clusters, the process can
be visualized as a dendrogram. By choosing some
cutoff-value in this dendrogram, we obtain the final clustering.

%
%
\subsection{Clustering: The Leiden algorithm}

In contrast to $k$-medoids or complete linkage clustering, the Leiden
algorithm\cite{Traag19} is performed on a graph, i.e., we need to
represent the correlation matrix in terms of nodes and edges.  Here
each node describes a coordinate of the system and the edges between
the nodes describe how similar (i.e., correlated) they are. Based on
such a graph, the Leiden algorithm identifies communities of similar
coordinates by unraveling the graph's cluster structure through a
maximization of an objective function (see below). The procedure
represents an improvement of the popular Louvain
algorithm,\cite{Blondel08} as it introduces randomness in the
selection of the community assignment, which facilitates a global
optimal partitioning by exploring the partition space more
broadly. (See the supplementary information in Ref.~\citenum{Traag19}
for a detailed description.) Roughly speaking, the algorithm consists
of three steps that are performed iteratively until the objective
function is not more improved. (1) The nodes are moved to the communities that
yield the highest gain of the objective function. (2) To improve the quality of
the partitioning, the communities of step 1 may be split into multiple
subcommunities. (3) The (sub)communities of step 2 now become nodes, thus
achieving a coarse graining of the graph.

A widely-used objective function $\Phi$ in community detection is
modularity, which indicates the amount of cluster-like structures in a
graph.\cite{Newman04} That is, high values of modularity
suggest that dense communities of nodes exist, which strongly interact
internally but are only loosely connected to the rest of the
graph. Low values, on the other hand, indicate that the graph shows
only little order but may be regarded as randomly
wired network. Modularity can be defined as
\begin{align}
    \Phi_{\text{mod}} = \frac{1}{2m}\sum_c \left(e_c - \frac{k^2_c}{2m}\right)\,,
\end{align}
where the sum runs over all clusters $c$, $m$ denotes the total number
of edges in the graph, $e_c$ is the sum of edge weights within cluster
$c$, and $k_c$ represents the sum of the degrees (i.e., number of
connecting edges) of all nodes in $c$.
While the first term favors the formation of large clusters, the
penalty term $k_c^2/2m$ prevents the clusters from becoming too large,
as it represents the expected numbers of edges within each cluster in
a randomly rewired graph, where the number of nodes are identical, and
all nodes keep their degree.  However, this implies that every node
can be connected to every other node, which is very unlikely in large
graphs, since the vicinity of each node only covers a small part of
the graph. Moreover, the expected number of edges in such large graphs
become very small (in the order of $N^{-1}$, where $N$ denotes the
number of nodes in the graph), which is why even a single link between
small communities might be interpreted as a sign of strong correlation
that may cause the merging of the communities independently of
their features. Therefore, if the graph is large enough, this implies
that two perfectly homogeneous clusters that represent completely
different collective motions would be merged at some point.
Ultimately this issue represents a resolution limit for the
identification of small clusters in large graphs.\cite{Kumpula07}

To circumvent this problem, Traag et al.\cite{Traag11} introduced a
new objective function, referred to as
``constant Potts model'' (CPM). As for modularity, CPM can be derived from
the Potts model,\cite{Wu84} which is a generalization of the Ising
model. The objective function reads
\begin{align} \label{eq:CPM}
    \Phi_\textsc{cpm} = \sum_c \left[e_c -\gamma \binom{n_c}{2}\right]\,,
\end{align}
where $n_c$ denotes the number of nodes in cluster $c$ and the
binomial $\binom{n_c}{2} = (n_c^2-n_c)/2$ represents the number of
possible edges within $c$. Weighting this number by the resolution
parameter $\gamma$, Eq.~\eqref{eq:CPM} compares the sum of edges in
cluster $c$ (i.e., the total correlation within $c$) to the edge sum
of a cluster of the same size but with all coordinates being
$\gamma$-correlated. If cluster $c$ performs better, it contributes
positively, otherwise the partitioning is penalized. In this way,
$\gamma$ determines the minimal average correlation required within
clusters. We note, however, that individual correlations can also be
below $\gamma$ if the overall function benefits from it, which is in
fact an important advantage over methods like complete linkage
clustering. Hence, the resolution parameter $\gamma$ controls the
coarseness of the clustering: A high $\gamma$-value and therefore a high
resolution leads to many communities, while lower resolution means
less, but also larger and thus more heterogeneous communities.

\subsection{Optimal Clustering Strategy}
\label{sec:clustering_strategy}

\begin{figure}[ht]
    \includegraphics{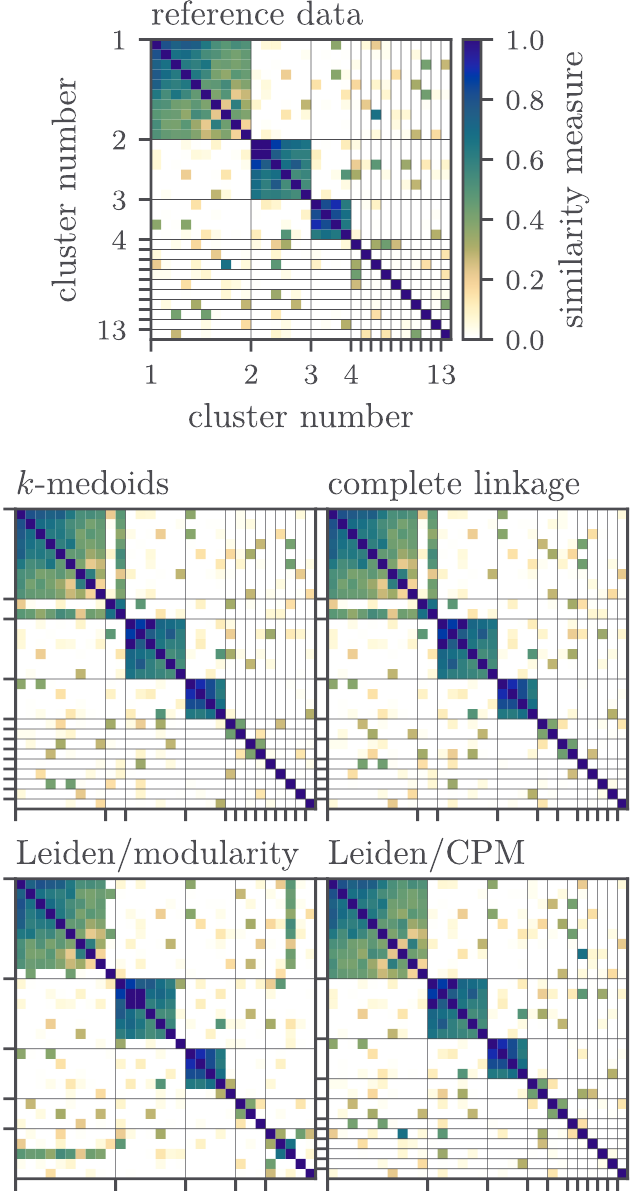}
    \caption{Simple model of a correlation matrix, consisting of three
      clusters of size 10, 6, and 4, as well as of 10 noise
      coordinates. Reference results are compared to the clustering
      methods $k$-medoids, complete linkage and the Leiden algorithm
      using modularity and CPM as objective functions.}
    \label{fig:community_methods}
\end{figure}

To compare the different clustering approaches introduced above, we
constructed as benchmark problem an artificial correlation matrix that
mimics the correlations found in protein systems. Inspired by our results
on HP35 and T4L below, we opted for a model that features three larger
clusters (containing various correlated coordinates) and multiple
mini-clusters reflecting noise coordinates. As a challenge, we also
included small residual correlations between the clusters (see SI Methods).
Figure~\ref{fig:community_methods} compares the resulting block-diagonal
correlation matrix to the outcome of the following clustering methods,
which were applied to the matrix with a random order of coordinates:
$k$-medoids, complete linkage and the Leiden algorithm using
modularity or CPM as objective function. To facilitate an unbiased
comparison, we spend some effort to optimize the parameters of each method.
That is, we used the silhouette method\cite{Rousseeuw87} to find initial
estimates, and determined the optimal value of the parameters via
V-measure\cite{Rosenberg07} which compares the clustering to the reference
results (Fig.~\SIFigCluster). This is relatively straightforward for
complete linkage and Leiden/CPM clustering that only require an
evident resolution parameter, but less so for $k$-medoids and
Leiden/modularity.

While all methods shown in Fig.~\ref{fig:community_methods} reproduce
the overall picture of main clusters and noise coordinates, they
differ in several possibly important aspects. That is, to successfully
identify collective protein motions, a clustering scheme should
fulfill the following requirements:
\begin{enumerate}
\item Completeness: All coordinates corresponding to a specific
  collective motion should be assigned to a single cluster. As a
  consequence, each cluster completely accounts for a specific
  collective motion.
\item Homogeneity: Each cluster should exclusively contain features
  which represent the same collective motion.\cite{Rosenberg07} This
  ensures that the clusters are as small as possible, such that noise
  coordinates can be reliably identified.
\end{enumerate}

In the light of these criteria, we notice that all algorithms except
Leiden/CPM do not satisfy the completeness criteria. For example, they
fail to correctly define cluster~1, which shows relatively strong
correlation to noise coordinates. Showing sporadic
correlation with coordinates that describe collective motion, noise
coordinates can lead to the formation of a new cluster, thus impeding
the completeness of larger clusters describing collective motion.
This is particularly an issue for $k$-medoids and
complete linkage clustering, which rely on greedy decision-making.
The Leiden algorithm using modularity is plagued by similar problems,
because it is based on a $k$-nearest neighbor graph that focuses
on the local neighborhood.\cite{Kumpula07} As discussed above,
modularity effectively introduces a resolution limit for small
clusters, which hampers the resolution of the noise coordinates in
Fig.~\ref{fig:community_methods}.

On the other hand, the Leiden algorithm combined with CPM, henceforth
referred to as ``Leiden clustering'', is able to obtain complete
clusters by permitting of locally suboptimal decisions, which
facilitate to find the global maximum of the objective
function. Moreover, the approach only requires to determine the single
intuitive resolution parameter $\gamma$. While these virtues of
the Leiden clustering may have only a small effect in
Fig.~\ref{fig:community_methods}, they are shown to result in significant
differences for the protein models studied below.

\section{Applications} \label{sec:appl}

We are now in a position to apply the above established strategy
(i.e., Leiden clustering of the linear correlation matrix) to identify
collective motions and the underlying functional mechanism of selected
protein systems. We start with the characterization of the cooperative
open-closed motion of T4 lysozyme\cite{Ernst17} (T4L), which
demonstrates the capability of Leiden clustering to distinguish
functional coordinates from noise. Studying the folding of villin
headpiece (HP35), we show that Leiden clusters can be a valuable means
to characterize the folding pathways of the system. All correlation
measures and clustering methods introduced above are implemented in
the Python package MoSAIC (``Molecular Systems Automated
Identification of Cooperativity''), which adapts the scikit-learn
syntax\cite{Pedregosa11} and is available on our homepage.\cite{note2}

\subsection{Functional Dynamics of T4 Lysozyme}

\begin{figure}[ht!]
    \includegraphics{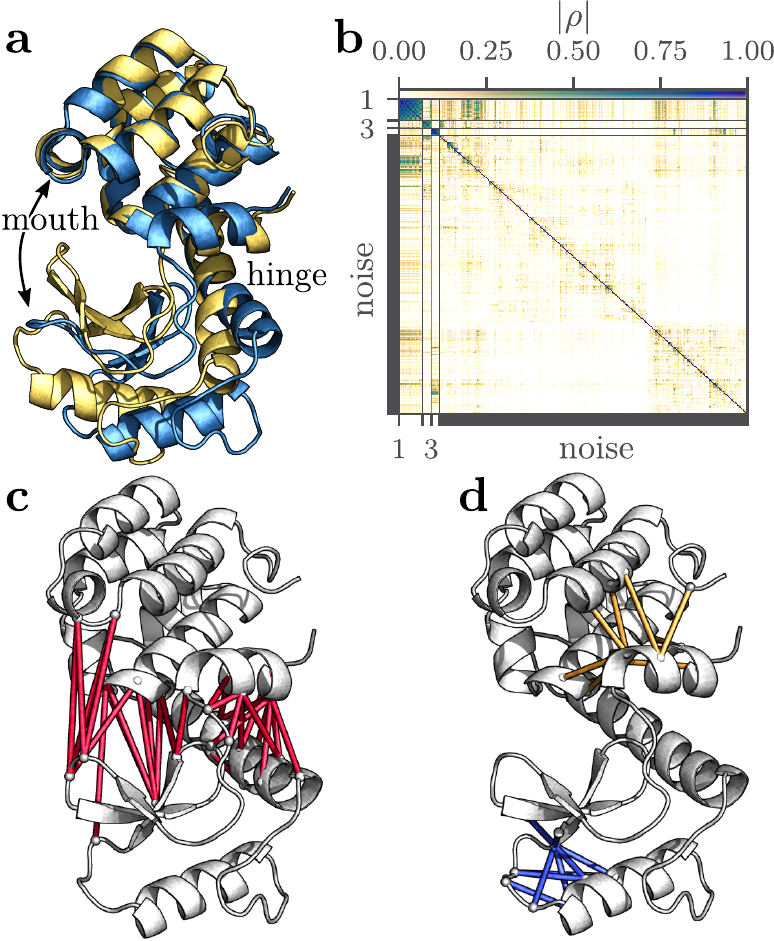}
    \caption{(a) Structure of the open (blue) and closed (yellow)
      states of T4L. The arrow indicates the Pac-Man-like open-closed
      transition of the mouth region, which is triggered by local
      motions in the hinge region. (b) Block-diagonalized correlation
      matrix of T4L, discriminating three clusters describing
      correlated motion from numerous noise coordinates. (c)
      Illustration of the interresidue contacts (red lines) contained
      in cluster~1 of the Leiden clustering, mediating the allosteric
      coupling between the distant mouth and hinge regions. (d)
      Interresidue contacts pertaining to cluster~2 (yellow) and
      cluster~3 (blue).}
    \label{fig:T4L}
\end{figure}

T4L is a 164-residue enzyme that performs an open-closed transition of
its two domains, which is triggered by local motions in the hinge
region\cite{Ernst17} (Fig.~\ref{fig:T4L}a). While the lifetimes of
the open and closed state are in the order of a few microseconds, the
transitions between these states occur on a nanosecond timescale, thus
indicating cooperative behavior. As discussed in previous work,
\cite{Hub09,Ernst17,Brandt18} the identification of reaction
coordinates underlying this cooperative process poses a challenge to
standard dimensionality reduction approaches.
Here we adopt the \SI{50}{\micro\second}-long all-atom MD simulation by
Ernst et al.,\cite{Ernst17} which was carried out using Gromacs 4.6.7
(Ref.~\citenum{Hess08}), the Amber ff99SB*-ILDN force field
\cite{Hornak06, Best09, Lindorff-Larsen10} and the TIP3P water
model.\cite{Jorgensen85} Assuming that a contact is formed if the
distance $d_{ij}$ between the closest non-hydrogen atoms of residues
$i$ and $j$ is shorter than \SI{4.5}{\angstrom} (Ref.~\citenum{Ernst15}),
$402$ native interresidue contacts were identified,\cite{Ernst17} and
the associated contact distances $d_{ij}$ were used to calculate the
linear correlation matrix.

When we perform Leiden clustering using a resolution parameter of
$\gamma = 0.5$ and assigning clusters containing $5$ or less
coordinates to noise, we obtain the block-diagonalized correlation
matrix shown in Fig.~\ref{fig:T4L}b. Interestingly, we find only three
main clusters representing correlated motion, while the majority of
coordinates are hardly correlated and thus distributed over the
remaining $\sim 300$ clusters.  To be specific, this means that the
average correlation within the main clusters is
$\langle |\rho| \rangle = 0.688$, while the mean residual correlation
between the three main clusters is $0.085$, and the residual correlation
between any two clusters (including the noise clusters) is on average
only $0.035$. This is because most intraprotein contacts of T4L are
quite stable and only fluctuate around their mean distance, while
contacts on the protein surface frequently form and break and hence
fluctuate randomly. In effect, all these coordinates represent noise
that should be discarded in a further analysis. We note that the other
clustering methods introduced above either fail to accurately
recognize the noise coordinates ($k$-medoids and Leiden/modularity) or
to properly identify the three main clusters ($k$-medoids and complete
linkage), see Fig.~\SIFigLys.

Having discarded the noise, we now turn to the first three clusters
that account for specific correlated motions of the system. In
particular, the 27 highly correlated contact distances of cluster~1
are found to describe the functional open-closed transition of T4L.
This is illustrated in Fig.~\ref{fig:T4L}c, which shows that these
distances span the space connecting the mouth and hinge regions, and
therefore reflect the allosteric coupling between these two distant
regions.\cite{Post22a}
Largely uncorrelated to cluster~1, clusters~2 and 3 account for other
correlated motions of T4L. As shown in Fig.~\ref{fig:T4L}d, cluster~2
describes a rocking motion involving the rearrangement of $\alpha_1$
and the N-terminal domain, while cluster~3 contains a twist-like
motion of the $\beta$-sheets and the close-by $\alpha_2$-helix. That
is, Leiden clustering is able to discriminate several independent
processes that occur at the same time. These motions were previously
found in higher principal components of various
PCAs,\cite{Hub09,Ernst17} and were therefore discussed as a part of
the functional dynamics. Being (almost) uncorrelated and thus not
involved in the open-closed transition of T4L, however, these
coordinates should not be included in the analysis of the open-closed
mechanism. Having said that, we note that in dynamical network theory
of allosteric communication, small residual correlations between
highly connected communities can nevertheless describe interactions
between several sub-processes of a protein.
\cite{Sethi09, McClendon09, Bowman12, Dokholyan16}

\subsection{Folding of Villin Headpiece}

\begin{figure}[ht!]
    \includegraphics{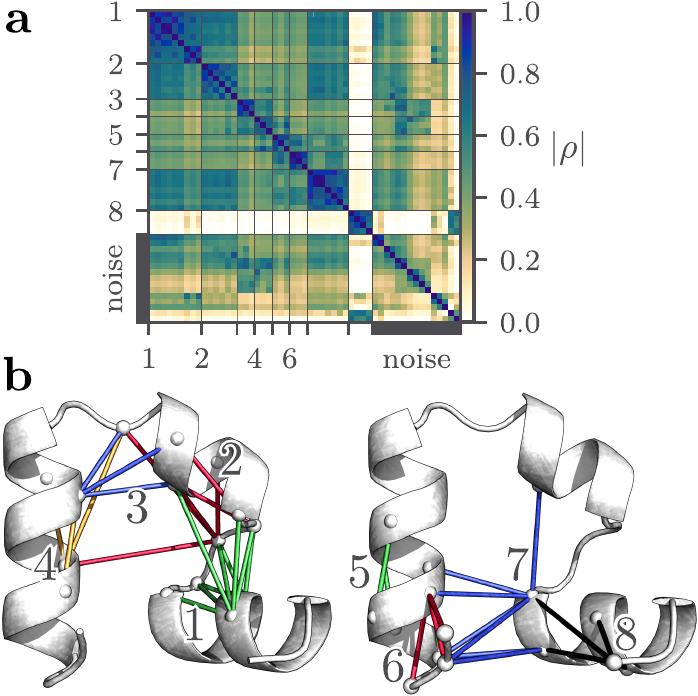}
    \caption{(a) Block-diagonalized correlation matrix of HP35,
      revealing seven main clusters, the overall uncorrelated
      cluster~8, and various noise coordinates. (b) Structure of HP35,
      indicating the contact distances included in clusters~1 to
      8. }
    \label{fig:HP35}
\end{figure}

HP35 is a 35-residues protein fragment that represents a popular model
of ultrafast protein folding.\cite{Kubelka08} It consists of a
hydrophobic core with three helices (residues 3--10, 14--19, and
22--32) that are connected via two unstructured loops, see Fig.\
\ref{fig:HP35}. Here we adopt a $\approx\SI{300}{\micro\second}$ long
all-atom MD trajectory of the fast folding Nle/Nle mutant at
$\SI{360}{\kelvin}$ of Piana et al.,\cite{Piana12} which reveals about
thirty folding events. Following Ref.~\citenum{Ernst15}, we use
the 53 native contacts of the crystal structure\cite{Kubelka06} (PDB 2f4k)
to construct the correlation matrix. Leiden clustering using a resolution
parameter of $\gamma=0.65$ yields the block-diagonal matrix shown in
Fig.~\ref{fig:HP35}a. It reveals 7 highly correlated main
clusters, the approximately overall uncorrelated cluster~8, and 10 weakly
correlated small clusters that are recognized as noise coordinates.
Although beeing divers in the partition of main and weakly correlated
clusters, the other clustering methods introduced above produce
roughly similar results, see Fig.~\SIFigVillin a.

To provide an intuitive picture of the Leiden partitioning of the
correlation matrix, Fig.~\ref{fig:HP35}b displays the individual
contact distances associated with the main clusters. First off, we see
that the coordinates of cluster~8 account for the motions of the
N-terminus relative to the $\alpha_1$-helix, which are completely
uncorrelated to the rest of the protein. As explained in the
Introduction, it is important to exclude such uncorrelated terminal
motion from the analysis. Dangling terminal residues may undergo
large-amplitude motion that is consequently recognized in the first
components of a PCA, although they are clearly not relevant for
folding. A related problem occurs if the uncorrelated motion exhibits
two-state behavior and performs transitions, e.g., between two
orientations of the terminus. As a consequence, the resulting number
of conformational states of the total system doubles trivially, which
unnecessarily complicates the analysis. As in the case of T4L, we thus
find that the identification and rejection of uncorrelated motions or
weakly correlated noise coordinates represents a crucial initial step
of a successful analysis.

Apart from the reduction of the dimension of the problem, Leiden
clustering may also support the interpretation of a considered
biomolecular process. To this end, we consider clusters~1 to 7, which
by design show high intracluster correlations
($\langle |\rho| \rangle = 0.818$), but also significant residual
correlations between the clusters (0.500 for main clusters only, 0.396
for all clusters), see Fig.~\ref{fig:HP35}a. Importantly, these
intercluster correlations indicate that all seven clusters describe
different aspects of the overall folding process. This is in contrast
to the findings for T4L, where the main clusters are almost
uncorrelated and thus describe independent processes.

With the exception of the small clusters~5 and (in part) 6 that account for
motions within a helix, the main clusters of HP35 consist of tertiary contacts
connecting two secondary structures (Fig.~\ref{fig:HP35}b). For
example, cluster~1 connects helices $\alpha_1$ and $\alpha_2$,
cluster~2 describes mostly correlations between the loops
$\alpha_1$-$\alpha_2$ and $\alpha_2$-$\alpha_3$, and cluster~7
connects the helices $\alpha_3$ and $\alpha_1$.
Due to high intracluster correlations, these contacts will preferably
form and brake in a concerted manner, such that we may assign a state
``1'' to a cluster if (most) of its contacts are formed, and a state
``0'' if not. Describing the complete protein by a product state
[e.g., (1110000) would mean that the contacts of the first three
clusters are formed, the others not], we can characterize the structures
of the folding trajectory in terms of this highly coarse grained state
description.\cite{Best13}
Since the clusters mostly refer to tertiary contacts, this state
partitioning is in contrast to a state definition via helicity where,
e.g., (ffu) means that the first two helices are folded and the
helix~3 not.\cite{Nagel20}

As a simple application of the above idea, we may ask which clusters
form first and which ones last in a successful folding event. By
analyzing the MD trajectory, we find that clusters~3 and 4 that
stabilize the connection of helices $\alpha_1$ and $\alpha_2$
typically fold first (Fig.~\SIFigVillin c). Most interestingly, we
learn that cluster~7 connecting the helices $\alpha_3$ and $\alpha_1$
forms last, and therefore represent the crucial step defining the
transition state of the folding process. The example demonstrates that
the coordinates defining the main clusters may give interesting first
hints on the mechanism of the considered process.

\section{Concluding remarks}

We have introduced a correlation analysis method for MD simulation
data, which aims to identify collective motions underlying functional
dynamics. Given some input coordinates such as interresidue distances
or backbone dihedral angles, the idea is to block-diagonalize the
corresponding correlation matrix and subsequently associate the
resulting clusters with functional motions or uncorrelated
noise. Notably, this strategy avoids possible bias due to presumed
functional observables\cite{Hub09} and conformational
states\cite{Brandt18} or variation principles that maximize
timescales.\cite{Scherer19}

To find the optimal algorithms for this workflow,
we have considered several linear and nonlinear correlation measures
and various clustering algorithms, which were implemented for a simple
and scalable use in the Python package MoSAIC.\cite{note2}
Interestingly, we have found that---at least for the considered
systems---the simple and well-established Pearson coefficient
describing linear correlation represents the best choice for this
purpose. This is because all considered nonlinear measures (i.e.,
various versions of a normalized mutual information) do not provide
essential new information (Fig.~\ref{fig:corr}a), while they focus
either on low or high values of the correlation (Fig.~\ref{fig:corr}b)
and require considerably higher ($\gtrsim 10^3$ times) numerical effort
(Fig.~\SIFigMI c).
Considering various clustering methods for matrix
block-diagonalization, we have shown that the Leiden community
detection algorithm\cite{Traag19} performs best for all considered
model systems. That is, Leiden clustering produces complete and
homogeneous clusters that can be clearly assigned to correlated
motions or noise (Fig.~\ref{fig:community_methods}).

Adopting the open-closed transition of T4L and the folding of HP35 as
representative examples, we have demonstrated the capability of Leiden
clustering to discriminate cooperative functional dynamics from
uncorrelated motions. In the case of T4L, \SI{89}{\%} of the 402 contact
coordinates were assigned to noise, which reflects the fact that these
contacts are either stable or fluctuate randomly (Fig.~\ref{fig:T4L}).
In the case of HP35, we found several highly correlated main clusters,
a large overall uncorrelated cluster, and various weakly correlated small
clusters that were recognized as noise (Fig.~\ref{fig:HP35}). Reflecting
large-amplitude motions of the N-terminus, the uncorrelated coordinates
should be excluded from the analysis. This is because they may be rated as
important by dimensionality reduction methods that maximize the variance
(such as PCA), although they are clearly not relevant for folding. Performing
Leiden clustering on the backbone dihedral angles of
HP35, we are similarly led to exclude the uncorrelated motions of
most of the $\phi$ angles, which (being slow but irrelevant) cause a
breakdown of TICA.\cite{Sittel18} Hence, we find that the
identification and rejection of uncorrelated motions or weakly
correlated noise coordinates represents a crucial initial step of a
successful dynamical analysis.

Apart from being a versatile feature selection scheme, Leiden
clustering may also facilitate a simple interpretation of the
considered biomolecular process. In the case of HP35, we obtained seven
main clusters that are directly associated with tertiary contacts
connecting secondary structures (Fig.~\ref{fig:HP35}). As contacts of
a cluster by design form and brake in a concerted manner, we can
employ the clusters as a coarse-grained state model to describe the
folding process.\cite{Best13} Considering the time evolution of the
clusters, for example, the model can be employed to illustrate the
folding pathways of the HP35.
In the case of T4L, Leiden clustering found three main clusters that
account for specific correlated motions of the system
(Fig.~\ref{fig:T4L}). While the 27 highly correlated contact distances of
the first cluster describe the functional open-closed of T4L, the
other two clusters account for other correlated motions of T4L and
should therefore not be included in the analysis.

Various extensions of the above correlation analysis are currently
under consideration. Following Tiwary and coworkers,\cite{Ravindra20}
for example, we aim to develop means to pick a few representative
coordinates from each cluster, in order to achieve a small number of
features describing the complete motion of a protein. Here, Leiden
clustering using the constant Potts model is particularly
advantageous, as we can simply control the extent of the coarse-graining
by setting the resolution parameter.
While we are generally reluctant to use Cartesian input coordinates
(apart from the separation problem\cite{note1}, the computation of the
correlation between non-scalar variables is notoriously difficult
\cite{Lange06}), it would be nevertheless interesting to compare the
outcomes of correlation analysis using contact distances and Cartesian
$\text{C}_\alpha$ coordinates. This is because they account differently for
non-trivial long-distance correlations, which are essential to
understand allostery in proteins. \cite{Wodak19} Although numerous
network models exist which aim to predict allosteric pathways,
\cite{Sethi09, McClendon09, Bowman12, Dokholyan16} the results
obtained from various formulations were found to differ significantly
even for simple model proteins such as PDZ domains.\cite{Lu16}

\begin{acknowledgement}

The authors thank Matthias Post, Steffen Wolf, Sofia Sartore and
Marius Lange for fruitful discussions, and D.~E.~Shaw Research for
sharing their trajectory of HP35. This work has been
supported by the Deutsche Forschungsgemeinschaft (DFG) via the
Research Unit FOR 5099 ``Reducing complexity of nonequilibrium''
(project No. 431945604). The authors acknowledge support by the
bwUniCluster computing initiative, the High Performance and Cloud
Computing Group at the Zentrum f\"ur Datenverarbeitung of the
University of T\"ubingen, the state of Baden-W\"urttemberg through
bwHPC and the DFG through grant No.~INST 37/935-1 FUGG.

\end{acknowledgement}

\begin{suppinfo}

The supplementary material contains details on the estimation of the
mutual information, the optimization of clustering parameters,
a comparison of different correlation measures and a comparison of T4L
and HP35 using all clustering methods discussed in the paper.

All discussed correlation measures and clustering methods are
implemented and freely available in the open-source software \emph{MoSAIC}
at \href{https://github.com/moldyn}{github.com/moldyn}.

\end{suppinfo}

\bibliography{/home/gs1002/Bib/stock, /home/gs1002/Bib/md, new}

\begin{mcitethebibliography}{75}
\bibitem[not()]{note1}
{Contrary to popular belief, the commonly employed rotational fit of the
  Cartesian trajectory to a reference structure cannot entirely remove the
  overall rotation of a flexible system. In particular, when applied to
  large-amplitude motions as found in protein folding, the residual overall
  rotation contained in the fitted trajectory may completely destroy the
  outcome of a subsequent dimenionality reduction\cite{Sittel14}}\relax
\mciteBstWouldAddEndPuncttrue
\mciteSetBstMidEndSepPunct{\mcitedefaultmidpunct}
{\mcitedefaultendpunct}{\mcitedefaultseppunct}\relax
\EndOfBibitem
\end{mcitethebibliography}

\end{document}


\section{Methods}

\subsection{Estimation of Mutual Information\label{SI:sec:MI_knn}}
In the main paper the mutual information is defined in
Eq. (4) via probability densities, that is, we rely on a
numerical robust estimation of the probability distributions. It is readily
shown that a simple histogram with fixed number of bins is not a reliable
estimate, as the number of bins affects the resulting estimate. Here, we
briefly introduce three different approaches to overcome the issue concerning
binning.

The easiest way of improving the density estimation is by using an optimal bin
width. To this end, we employ the Freedman-Diaconis rule\cite{Freedman81} with
an adjusted prefactor of 2.59 which defines the bin width $d$ as
\begin{align}
	d = 2.59 \frac{\text{IQR}(X)}{\sqrt[3]{N}} \label{SI:eq:freedman-diaconis}
\end{align}
where $\text{IQR}$ is the interquartile range of $X$, and $N$ is the number of
samples. This rule was designed to yield good results for normal distributions,
but struggles with distributions featuring fat tails.
For example, in the case of contact distances, we typically find distributions
with a single-sided fat tail. As a remedy, we may scale the bin width by the
following factor
\begin{align}
    d\to d \cdot \frac{\text{100th percentile}-\text{0th percentile}}{\text{85th percentile}-\text{15th percentile}}\,,
\end{align}
which yields closely matching estimates with a $k$-nearest neighbor ($k$-nn)
estimator.

Alternatively, we may estimate the probability densities via kernel density
estimation (KDE), and employ Scott's rule\cite{Bashtannyk01} for the estimation
of the optimal bandwidth. However, this can be problematic since the probability
distributions for multi-modal distribution can easily be over-smoothed.
To overcome this problem, Kraskov proposed a mutual information estimator
based on the $k$-nn distribution,\cite{Kraskov04}
\begin{align}
    I(X, Y)&= \psi(N) + \psi(k)
        - \frac{1}{N}\sum\limits_{i=1}^{N}\left[\psi(n_{x,i} + 1) + \psi(n_{y,i}+1)\right]\,,
    \label{SI:eq:MI_knn_estimate}
\end{align}
where $\psi$ denotes the digamma function and $n_{x,i}$ represents the number of
points $j$ which lay within $||x_i - x_j || < \varepsilon_i^k$.
For each point $i$, the cut-off value $\varepsilon_i^k$ is chosen as the maximal
$k$-nearest neighbor distance in $x$ or $y$
($\varepsilon_i^k=\max\{\varepsilon_{x,i}^k, \varepsilon_{y,i}^k\}$).

\begin{figure}[ht!]
    \centering
    \includegraphics{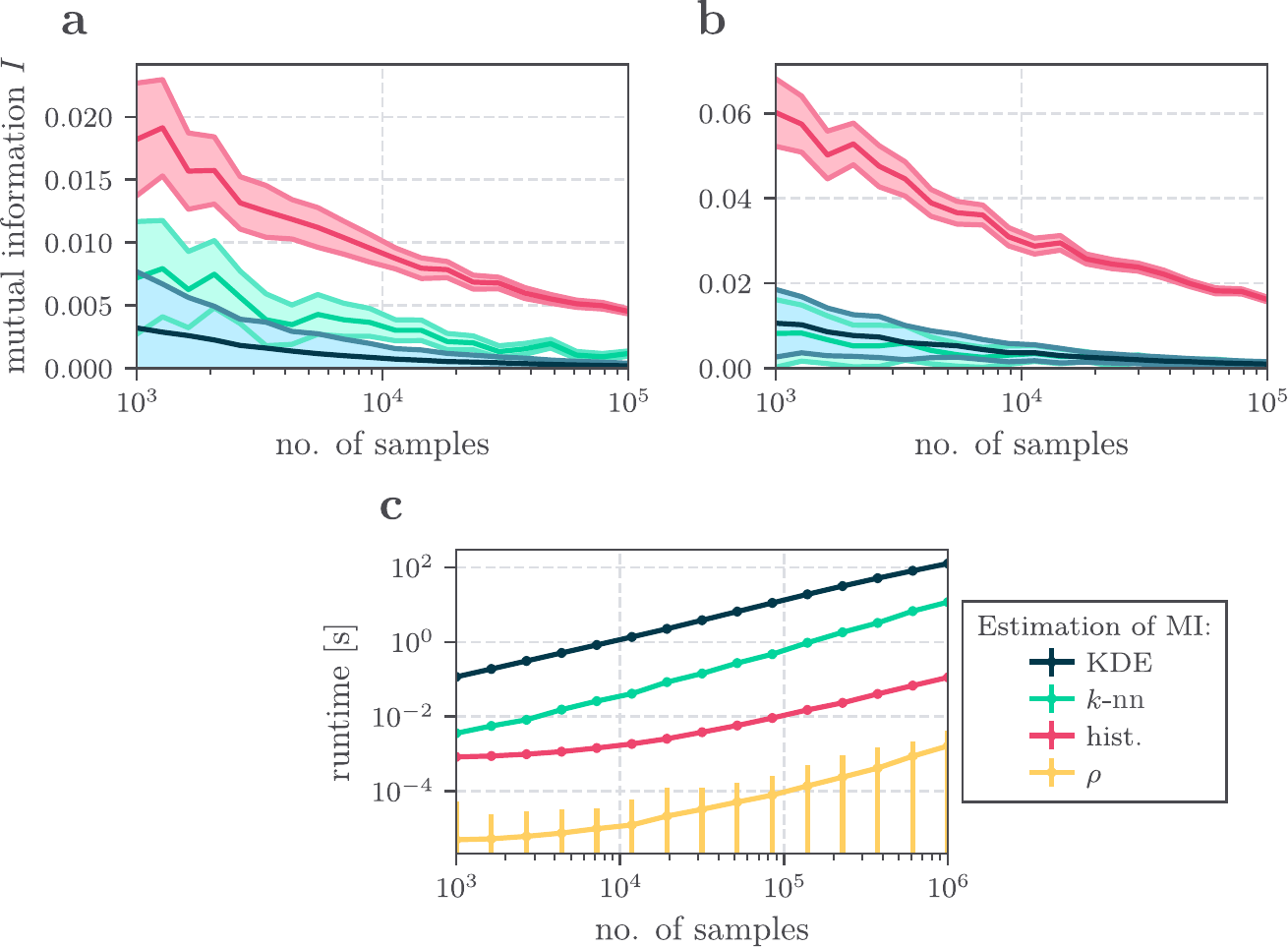}
    \caption{Considering two independent (a) uniform distributions and (b)
    normal distributions, the mutual information is calculated via a
    histogram (red), a nonparametric method based on entropy estimation
    from $k$-nearest neighbors distances (green), and by a kernel density
    estimation (dark green).
    Mean and standard deviation are estimated from 100 independent runs.  (c)
    Runtime comparison of various correlation estimators for two normally
    distributed variables, using a single core of an
    Intel\textsuperscript{\textregistered} Core\texttrademark{} i9-10900
    CPU.
    \label{fig:SI:MI}}
\end{figure}

In the following we are interested in the error of the mutual information
estimation. Therefore, we use independent random variables which have a
vanishing mutual information $I=0$.
In Fig.~\ref{fig:SI:MI} we generated (a) two uniformly independent
random variables bound to $[0, 1)$, and (b) two normally distributed random
variables with zero mean and standard deviation $\sigma=1$.
The results clearly show that the mutual information converges much faster for
$k$-nn and KDE than for a simple histogram with optimized bin width.
In terms of absolute values however, the ansatz of dynamically selecting the
number of bins according to Eq.~\eqref{SI:eq:freedman-diaconis} is sufficiently
precise.

On the other hand, there is the problem of runtimes. While the histogram
approach has the largest error, it is also by far the fastest. In
Fig.~\ref{fig:SI:MI}c, we compare the runtimes of the three methods discussed
including the linear correlation. To make the results as comparable as
possible, we have limited the calculations to a single core of an
Intel\textsuperscript{\textregistered} Core\texttrademark{} i9-10900 CPU and
two normal distributed variables. The total runtime scales with the number of
features $M$ by $\frac{1}{2} M (M - 1)$. Hence, only linear correlation is
feasible when considering numerous features.

\subsection{Optimization of clustering parameters\label{SI:Sec:V_measure}}
\begin{figure}
    \centering
    \includegraphics{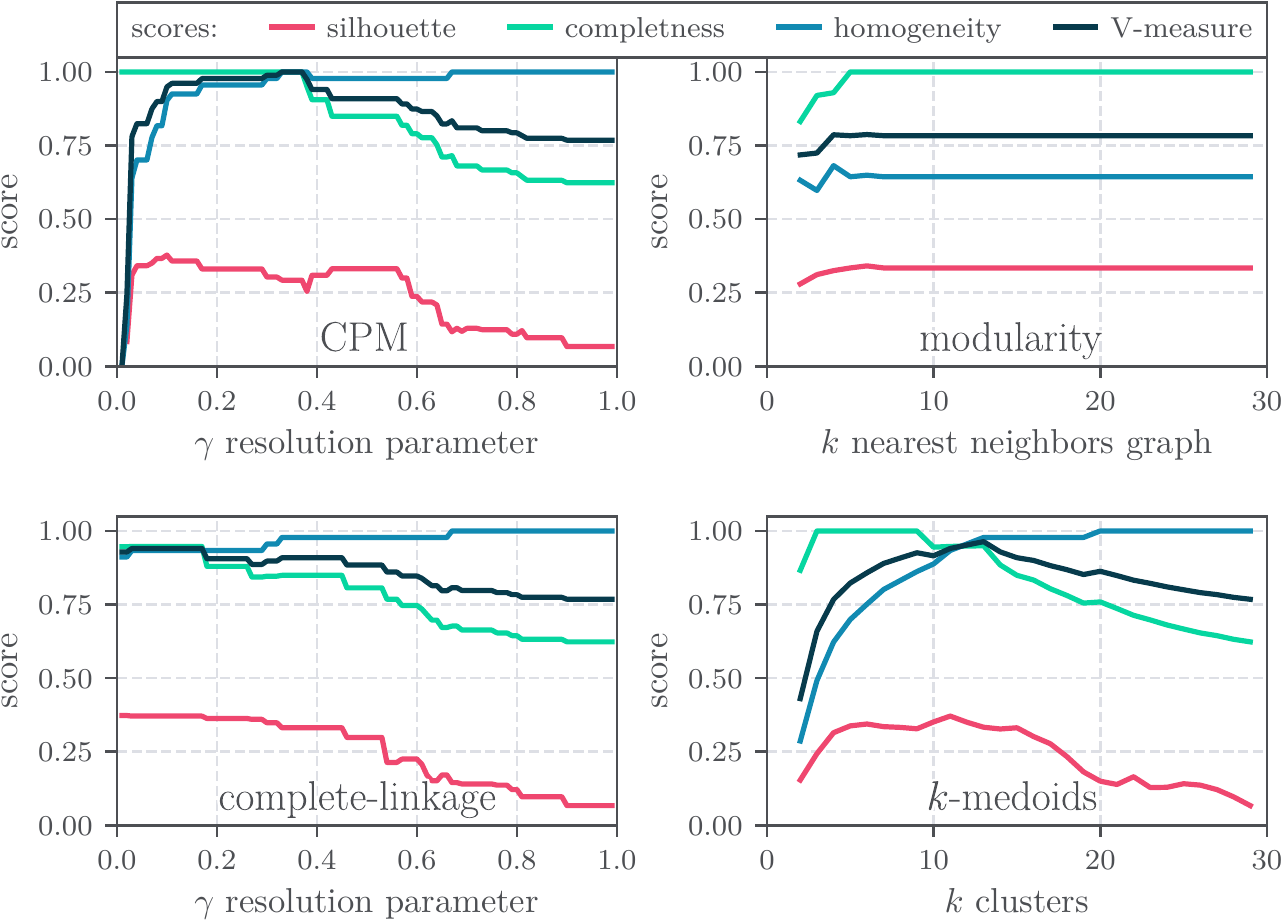}
    \caption{Comparison of the performance of Leiden/CPM, Leiden/modularity,
    complete-linkage and $k$-medoids clustering, using the silhouette
    method [Eq.~\eqref{eq:SI:silhouette}] and the V-measure
    [Eq.~\eqref{eq:vscore}].
    \label{fig:SI:v_silhouette_measure}}
\end{figure}

In the case of the simple model of a correlation matrix
(Sec. 3.2), we by design know the ground truth of the
clusters. The matrix is available at our homepage.\cite{note2}
Hence, we can calculate the V-measure according to\cite{Rosenberg07}
\begin{align}
    V & = \frac{2hc}{h + c},
    \label{eq:vscore}
    \intertext{where $c$ is completeness and $h$ homogeneity, defined as}
    c & =   \left\{
                \begin{array}{ll}
                    1 & \text{if } H(K, C) =  0 \\
                    1 - \frac{H(C|K)}{H(K)} & \text{else} \\
                \end{array}
            \right.
    \intertext{and}
    h & =   \left\{
                \begin{array}{ll}
                    1 & \text{if } H(C, K) =  0 \\
                    1 - \frac{H(C|K)}{H(C)} & \text{else}\;.\\
                \end{array}
            \right.
\end{align}
Here $C$ is the set of ground truth (compare Fig. 2a
in the main paper), and $K$ is the set of the resulting clusters.
$H(K, C)$ is the joint entropy and $H(K|C)$ the conditional entropy of
the resulting cluster partition given the ground truth.\cite{Rosenberg07}
The results for all possible clustering parameters can be found in
Fig.~\ref{fig:SI:v_silhouette_measure}.

Usually however, we do not know the ground truth. In this case, we can use the
so-called silhouette method as a heuristic.\cite{Rousseeuw87} For each feature
$i \in C_I$ belonging to cluster $C_I$ we can define the mean distance between
$i$ and all other features belonging to the same cluster by
\begin{align}
    a_i = \frac{1}{|C_I| - 1} \sum_{\substack{j\in C_I\\ j\neq i}} d_{ij}\;,
    \intertext{and its average distance to its nearest neighbor cluster}
    b_i = \min_{J\neq I}\frac{1}{|C_J|} \sum_{j\in C_J} d_{ij}\;.
\end{align}
Therewith, the silhouette coefficient can be defined for all $M$ features by
\begin{align}
    SC &= \langle s_i \rangle_{i\in M}\;,\label{eq:SI:silhouette}
    \intertext{where the contribution of each feature $i$ is defined by}
    s_i &= \begin{cases}
        1 - \frac{a_i}{b_i} & \text{if}\; a_i < b_i\\
        0 & \text{if}\; a_i = b_i\\
        \frac{b_i}{a_i} - 1 & \text{if}\; a_i > b_i\;.
    \end{cases}
\end{align}
In Fig.~\ref{fig:SI:v_silhouette_measure} we study the effect of changing the
clustering parameters on the silhouette coefficient. Comparing it to the
previous results of V-measure we find similar results. E.g., when the latter is
minimal also the silhouette coefficient is minimal. Nevertheless, we find a
slight shift of the maxima. Hence, we advise to use the silhouette method only
as a guide to find a first estimate on the clustering parameter, but not
necessarily as a way to determine the final clustering parameters.

\section{Results}
\subsection{Linear vs. Nonlinear Correlation Measures}
\begin{figure}[ht!]
    \centering
    \includegraphics[width=\textwidth]{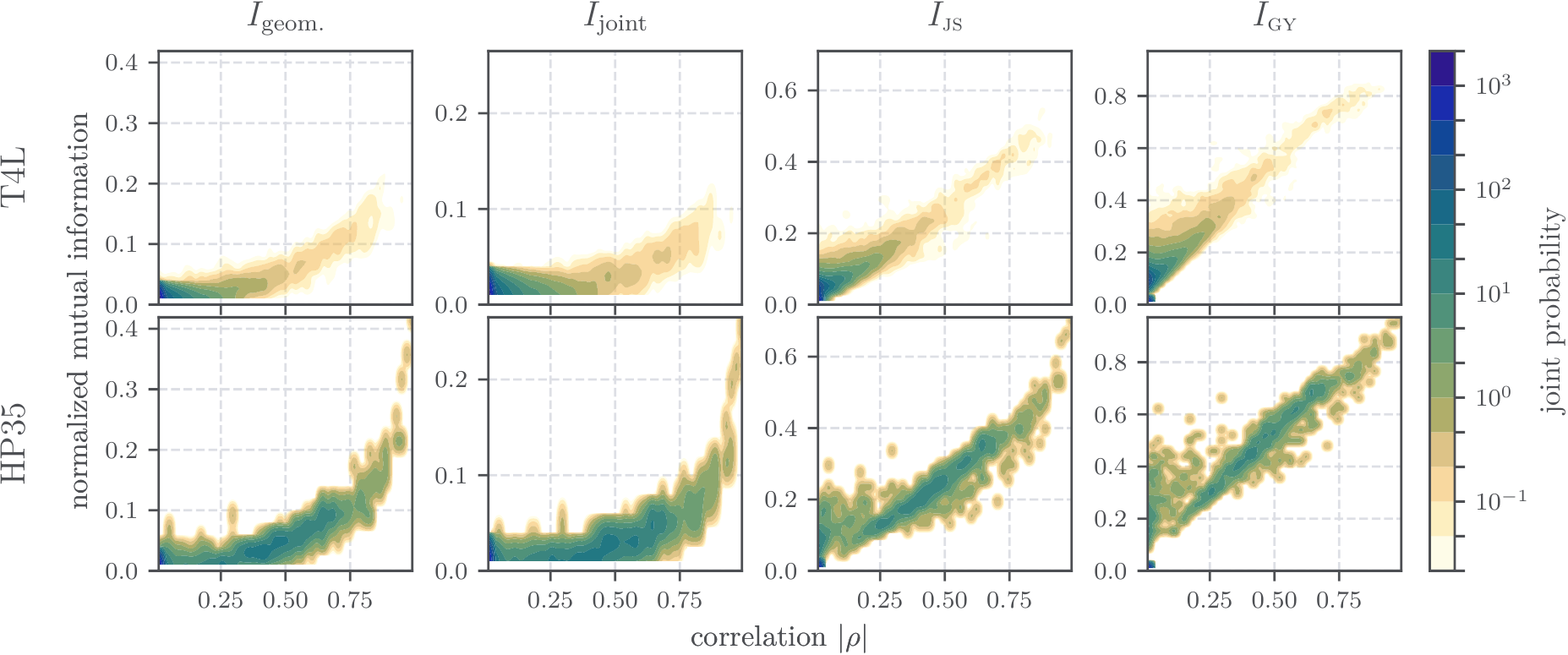}
    \caption{Comparison of the nonlinear correlations $I_\text{geom.}$,
    $I_\text{joint}$, $I_\textsc{js}$, and $I_\textsc{gy}$ to the absolute
    Pearson coefficient $|\rho|$ for HP35 and,T4L.
    \label{fig:SI:similarity_comparison}}
\end{figure}

\clearpage
\subsection{Clustering of T4L}
\begin{figure}[ht!]
    \centering
    \includegraphics{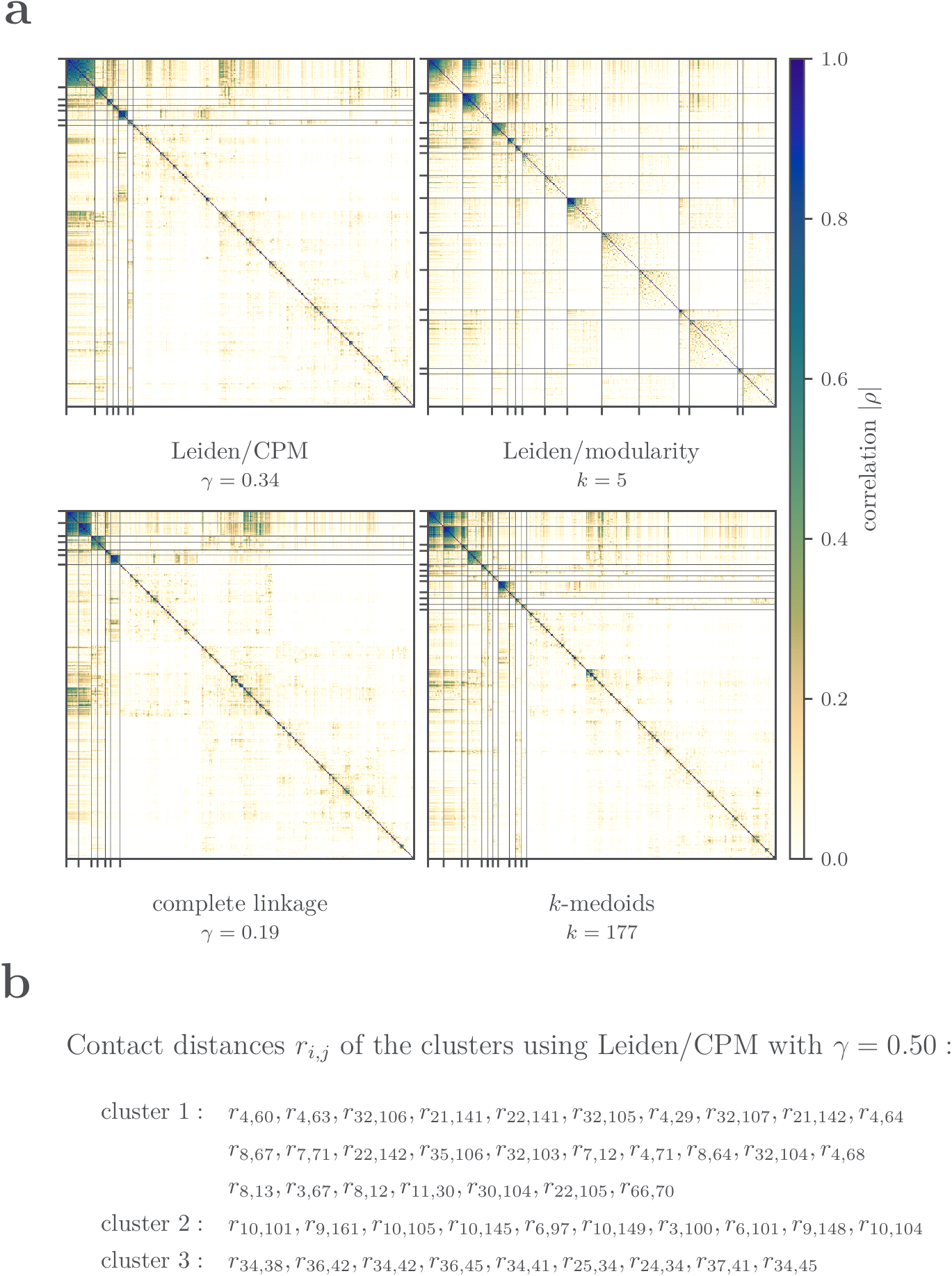}
    \caption{(a) Clustering the 402 contact distances of T4L using Leiden/CPM,
    Leiden/modularity, complete-linkage and $k$-medoids with optimized parameters
    according to silhouette score. (b) Description of all clusters in
    Fig. 3 with more than 5 coordinates.
    \label{fig:SI:T4L}}
\end{figure}

\clearpage
\subsection{Clustering of HP35}
\begin{figure}[ht!]
    \centering
    \includegraphics[width=0.98\textwidth]{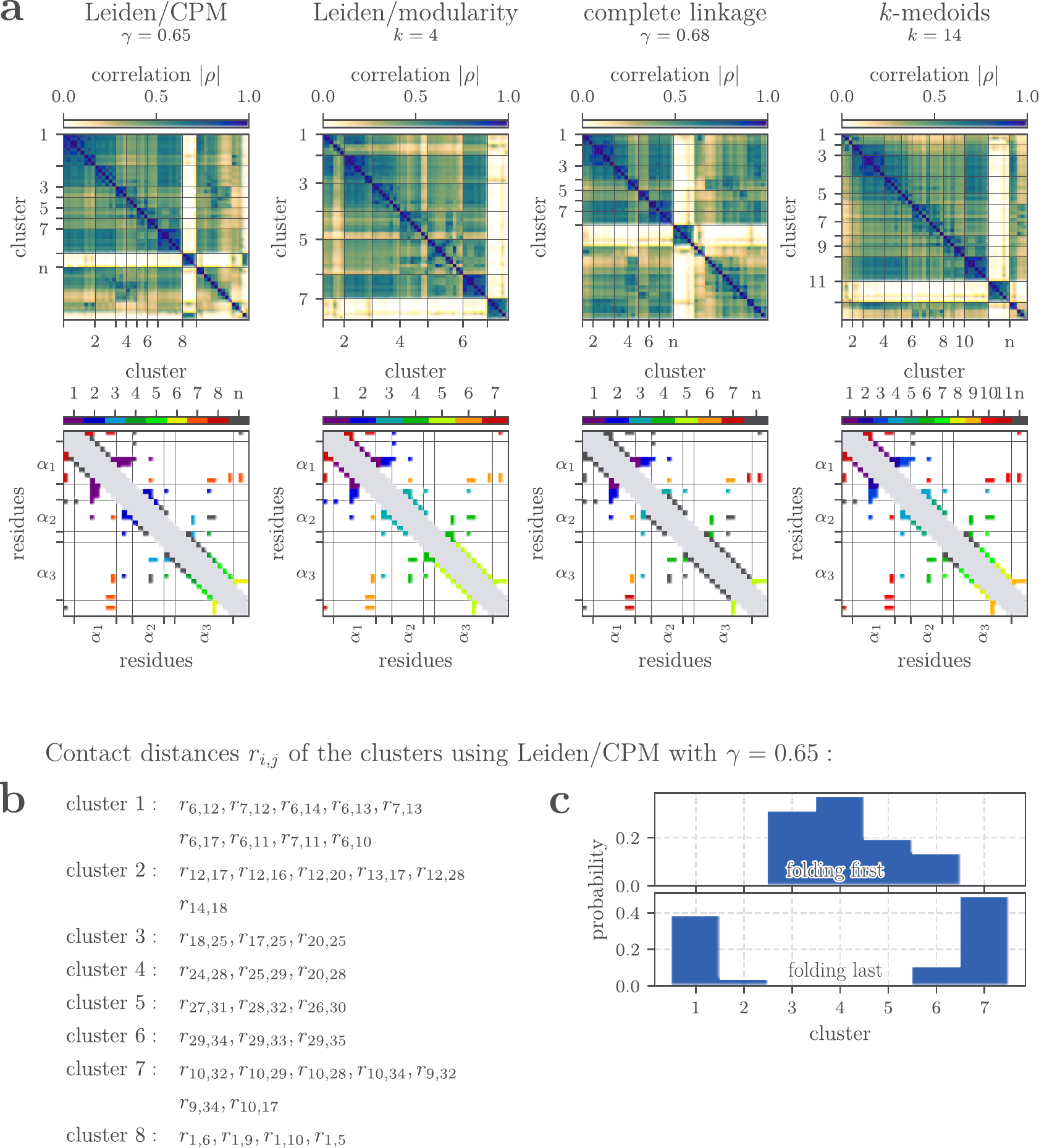}
    \caption{(a) Clustering the 53 contact distances of HP35 using Leiden/CPM,
    Leiden/modularity, complete-linkage and $k$-medoids with optimized
    parameters according to silhouette score. Where the resulting clusters are
    visualized by their corresponding (top) correlation matrix and (bottom)
    contact map.
    (b) Description of all clusters in Fig. 4 with more than one
    coordinate. (c) Probability that the clusters in Fig. 4 are
    folding first/last within one of the 31 folding events identified by Piana
    et al.\cite{Piana12}.
    \label{fig:SI:HP35}}
\end{figure}
\clearpage

\bibliography{/home/gs1002/Bib/stock, /home/gs1002/Bib/md, new}